\documentclass{svproc}
\usepackage{graphicx,subfigure}
\usepackage{amsfonts}
\usepackage{amsmath}
\usepackage{url}

\begin{document}
\mainmatter              
\title{Melody Classifier with Stacked-LSTM}
\titlerunning{Stacked-LSTM}  
%
\author{You Li\inst{1} \and Zhuowen Lin\inst{2}}
\authorrunning{You Li et al.} 
%
\tocauthor{You Li, and Zhuowen Lin}
\institute{Steinhardt School of Culture, Education, and Human Development, New York University, New York City, United States
\and School of Electrical and Computer Engineering, Georgia Institute of Technology, Atlanta, United States}

\maketitle              

\begin{abstract}
Attempts to use generative models for music generation have been common in recent years, and some of them have achieved good results. Pieces generated by some of these models are almost indistinguishable from those being composed by human composers. However, the research on the evaluation system for machine-generated music is still at a relatively early stage, and there is no uniform standard for such tasks. This paper proposes a stacked-LSTM binary classifier based on a language model, which can be used to distinguish the human composer's work from the machine-generated melody by learning the MIDI file's pitch, position, and duration.
\keywords{Melody classification, Language Model, LSTM}
\end{abstract}
\section{Introduction}
    Significant progress in the realm of music generative models has been made in the last few years. To capture the salient characteristics of music data, and to generate new music samples that are indistinguishable from the true data are two main goals of generative modeling. Several deep neural network architectures for music generation have been proposed, like WaveNet\cite{oord2016wavenet}, MuseGAN\cite{dong2017musegan}, and Jukebox\cite{dhariwal2020jukebox}, and have brought exciting innovations.
	
	However, the development of music generative models blurs the borderline between human music and artificial intelligence (AI) generated music. Legal issues may arise in music industry when high-accuracy generative models are utilized intentionally to conduct music plagiarism. Therefore, a model that can be used to do classification on human-composed music and AI-composed music is of great importance.
	
	In this paper, as part of the AI Composition Recognition Competition held by Conference on Sound and Music Technology, we introduce the implementation details of the competition, including dataset preparation and the stacked-LSTM model we used to distinguish AI-generated music from human-composed music.

\section{Implementation}
	\subsection{Dataset Preparation}
	The development dataset provided by the committee was used as training data in label 0, because it only contains MIDI files generated by AI algorithms. In order to use our LSTM model, which was a supervised learning model, we used a dataset that entirely consisted of melodies written by human musicians to form our training data in label 1. It was crawled, collected, and posted five years ago on a Reddit post by the user "midi\_man" \cite{reddit_data} and it is still available and free to use by the date of submission of this paper. Training data in label 0 and 1 were used collectively to train our neural network model, and we split 40\% of them for validation. The evaluation dataset provided by the committee was used as test data to test our trained model and generate final results. 
	
	The training data in label 0 were monophonic melodies with the length of 8 bars each, so we manually cut MIDI files in label 1 with Digital Audio Workstation (DAW) to let them have the same structure, and further, quantized them to make the MIDI events on the grid. Because the MIDI events in training data in label 0 and test data were all off the grid, in order to eliminate the possibility that the neural network model would learn to classify the test data purely with beats of MIDI events, we wrote a function in our MIDI parser to stretch or quantize the training data in label 0 and test data according to their BPMs.
	
	The final composition of training dataset was 6000 pieces in label 0 and 1000 pieces in label 1, covering several genres including classical, pop, and electronic.
	
	\subsection{Features and Representation}
	Usually, MIDI files contain massive information, including but not limited to notes and MIDI control change messages. As for notes, the most critical features are pitch, position, duration, and velocity. Unfortunately, in many MIDI files, the velocity value is set fixed, so we cannot use it effectively.
	
	The features we chose were the pitch and duration of notes extracted from the melody and their corresponding position (beat) in each bar. The following figure shows the feature map within a single bar, in which the first column represents the pitch, and the rest are position and duration, respectively. It is worth noting that the position ‘0.0’ refers to the first downbeat of a measure in music theory, and ‘1.0’ represents the duration of a quarter note.
	
    \begin{figure}[htb]
		\centering
		$ mapping = \begin{bmatrix}
		'C4' ,& '0.0' ,& '1.0'\\ 
        'E4' ,& '1.0' ,& '0.5'\\ 
        'G4' ,& '1.5' ,& '0.5'\\ 
        'C5' ,& '2.0' ,& '2.0' 
		\end{bmatrix}$
		\caption{Feature mapping in a bar.}
		\label{fig:figure1}
	\end{figure}
	
	To meet the LSTM input requirements, we performed one-hot encoding on the three features separately and stacked them along the 0-axis to form a sparse matrix. In this way, the weights of the features could be updated at the same time. Also, LSTM requires the input sequence to have a fixed length, so we padded the input sequence according to the length of the most extended sequence in the training set.

    \subsection{Model}
	As for the model, we chose to use a stacked-LSTM model. A similar structure was first introduced by Graves et al. \cite{graves2013speech} in 2013. Their speech recognition experiments found that deeper recurrent neural networks could significantly improve the model’s performance in dealing with sequence inputs.
	
	Our model contained two LSTM layers. The first one had 64 units. Each time step had a hidden state output for every single LSTM unit in this layer, which would be used as the input of the second LSTM layer. The second LSTM layer had eight units. In this layer, we only took the final output of the sequence. Finally, we used a fully connected layer to obtain the final classification results. We added two dropout layers between the LSTM layers and the fully connected layer to prevent overfitting, with a dropout rate of 0.4. One of the advantages of this model are, due to its deep neural network structure, it can better learn more abstract and global features from the sequence. The following figure shows the detailed structure of our model.
	
	\begin{figure}[htb]
		\centering
		\includegraphics[scale=0.5]{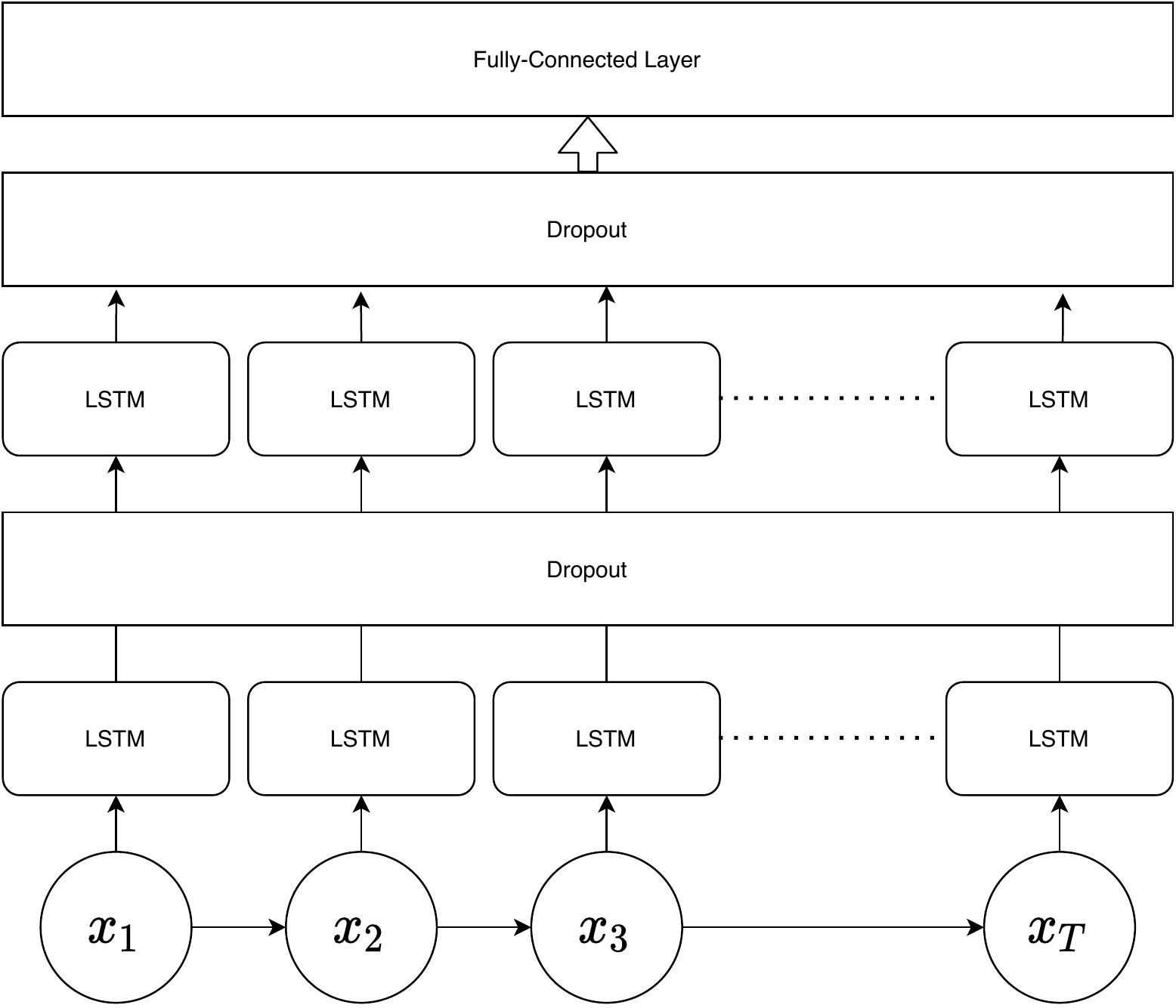}
		\caption{Architecture of the model}
		\label{fig:figure}
	\end{figure}

	We also tried bidirectional LSTMs, the model’s performance on the validation set did not improve significantly, but we still included the prediction in the final results.

\section{Conclusion and Further Work}
	We proposed and trained a stacked-LSTM neural network model to do classification on human-composed music and AI-composed music. It is submitted along with the classification result to the competition committee to evaluate its performance. 
	
	Future work can be done in finding a faster way to increase the amount and diversity of real data suitable for model training. 

\bibliographystyle{spmpsci}     
\bibliography{mybib}

\begin{thebibliography}{1}
\providecommand{\url}[1]{{#1}}
\providecommand{\urlprefix}{URL }
\expandafter\ifx\csname urlstyle\endcsname\relax
  \providecommand{\doi}[1]{DOI~\discretionary{}{}{}#1}\else
  \providecommand{\doi}{DOI~\discretionary{}{}{}\begingroup
  \urlstyle{rm}\Url}\fi

\bibitem{dhariwal2020jukebox}
Dhariwal, P., Jun, H., Payne, C., Kim, J.W., Radford, A., Sutskever, I.:
  Jukebox: A generative model for music.
\newblock arXiv preprint arXiv:2005.00341  (2020)

\bibitem{dong2017musegan}
Dong, H.W., Hsiao, W.Y., Yang, L.C., Yang, Y.H.: Musegan: Multi-track
  sequential generative adversarial networks for symbolic music generation and
  accompaniment.
\newblock arXiv preprint arXiv:1709.06298  (2017)

\bibitem{graves2013speech}
Graves, A., Mohamed, A.r., Hinton, G.: Speech recognition with deep recurrent
  neural networks.
\newblock In: 2013 IEEE international conference on acoustics, speech and
  signal processing, pp. 6645--6649. IEEE (2013)

\bibitem{reddit_data}
midi\_man: The largest midi collection on the internet, collected and sorted
  diligently by yours truly. (2015).
\newblock
  \urlprefix\url{https://www.reddit.com/r/WeAreTheMusicMakers/comments/3ajwe4/the_largest_midi_collection_on_the_internet/}.
\newblock [Online; accessed 7-August-2020]

\bibitem{oord2016wavenet}
Oord, A.v.d., Dieleman, S., Zen, H., Simonyan, K., Vinyals, O., Graves, A.,
  Kalchbrenner, N., Senior, A., Kavukcuoglu, K.: Wavenet: A generative model
  for raw audio.
\newblock arXiv preprint arXiv:1609.03499  (2016)

\end{thebibliography}

\end{document}